\documentclass[journal]{IEEEtran}
\usepackage{array}
\usepackage[tight,footnotesize]{subfigure}
\usepackage{bm}
\usepackage{amsfonts}
\usepackage{amsmath}
\usepackage{epsfig,float}
\usepackage{epstopdf}
\usepackage{graphicx}

\usepackage{enumitem}

\def\bE{\mathbb{E}}

\def\cR{{\cal R}}

\def\be{\begin{equation}}
\def\ee{\end{equation}}
\def\ba{\begin{align}}
\def\nn{\nonumber\\}
\def\ber{\begin{eqnarray}}
\def\eer{\end{eqnarray}}
\def\defeq{\stackrel{\rm def}{=}}
\def\maxm{\mathop{\rm max}_{\scriptstyle m}}

\def\minm{\mathop{\rm min}_{\scriptstyle m}}
\def\MPE{P_e^{(min)}}
\newcommand{\argmax}[1]{\underset{#1}{\operatorname{arg}\operatorname{max}}\;}

\begin{document}
\title{\bf Bayesian Error Based Sequences of Mutual Information Bounds}
\author{Sudhakar~Prasad,~\IEEEmembership{Fellow,~OSA}%
\thanks{The author is with the Department of Physics
and Astronomy, University of New Mexico, Albuquerque, New Mexico 87131}
\thanks{Manuscript submitted to IEEE Trans. Inform. Th., 22 Sep 2014}}

\maketitle
\begin{abstract}
The inverse relation between mutual information (MI) and Bayesian error is sharpened
by deriving finite sequences of upper and lower bounds on MI in terms of the minimum probability of error (MPE)
and related Bayesian quantities. The well known Fano upper bound and Feder-Merhav lower bound on equivocation
are tightened by including a succession of posterior probabilities starting at the largest, which directly controls
the MPE, and proceeding to successively lower ones. A number of other interesting results
are also derived, including a sequence of upper bounds on the MPE
in terms of a previously introduced sequence of generalized posterior distributions.
The tightness of the various bounds is illustrated for a simple application of
joint spatial localization and spectral typing of a point source.
\end{abstract}
\IEEEpeerreviewmaketitle

\section{Introduction}\label{sec:intro}

Statistical information theory \cite{Shannon48,CT06} is an essential quantitative tool for
modern signal processing, computation, coding, and communication systems.
Its core philosophy hinges on the notions of information potential and
the ability of systems to encode, transmit and decode information
about one or more input parameters.

Since information must improve with decreasing error,
MI is expected to have an inverse relationship with fundamental Bayesian error
metrics, such as the minimum mean squared error (MMSE) and the minimum probability
of error (MPE) \cite{VBT13}. Two lower bounds on MI, one involving the MPE and another the MMSE,
were derived over forty years ago by Fano \cite{CT06} and Seidler \cite{Se74}, respectively. While there has been
some progress made in deriving useful upper bounds \cite{FM94,SV06} as well in the last twenty years, there
is still need for tighter bounds than are currently available.
The relationship between the MPE and MMSE for highly sensitive detectors has been clarified in a recent paper
by the author \cite{SP12}.

The matter of MI bounds has also received attention
because of the essential relation of MI to code redundancy
in universal source coding \cite{CT06,D73} and to Bayesian risk in multi-hypothesis testing and computational learning
as elucidated by Haussler and Opper \cite{HO95}. In the latter work, the authors
derived a lower bound and an upper bound on MI involving double expectations of
exponentiated Hellinger and Kullback-Leibler distances between two conditional data PDFs, respectively,
These bounds, while tight in the asymptotic limit, are expected to be loose, however,
when such a limit cannot be taken.

A related topic of bounding the MPE for multi-hypothesis testing
in the Bayesian setting has received much attention for two main reasons, one being the need for computational
efficiency and the other being its usefulness in developing and proving certain
asymptotic results in coding theory. Bounds of this type include the Fano lower bound (FB) involving
the equivocation entropy (EE) \cite{CT06} and its generalization for nonuniform priors \cite{HV94}
and countably infinite alphabets \cite{HV10}, the Shannon bound \cite{Sh57},
the Poor-Verdu bound \cite{PV95}, and the generalized Poor-Verdu
bound \cite{CA12}. Of these, only the FB and its generalizations yield bounds on the EE and thus on MI,
which is the subject of main interest here.

The converse problem of bounding the EE in terms of the MPE has been elucidated by
Feder and Merhav \cite{FM94}. They derived rigorous convex and non-convex lower bounds as functions
of the MPE alone, and showed their usefulness in formulating and proving the
channel coding theorem by upper-bounding the EE, rather than MPE,
as a function of the codeword length and transmission rate.
Their EE bounds, while being the tightest possible when using the MPE alone,
will be shown, however, to be rather loose for an application considered here.
One purpose of the present paper is to tighten these lower bounds on EE by introducing secondary Bayesian measures
of error that successively refine these bounds upwards while being amenable to physical interpretation.
We also derive, in terms of the same secondary error measures, a finite sequence of upper bounds on EE
that tighten the Fano bound as well, thus furnishing two sequences of upper and lower bounds
on the EE - and thus on MI - that tighten the existing ones.
We also present a different upper bound on MI, closely related to the Feder-Merhav (FM) bound, that is tighter than all others, but has no
simple physical interpretation.

As part of our review and elaboration of the FM lower bound on the EE,
we shall obtain a number of alternative derivations
for it. We shall use one of them to also upper-bound
the MPE by means of a sequence of bounds that increasingly tighten that furnished by the FM bound.
These bounds, as we shall see, involve the family of generalized or tilted posterior probability mass functions
(PMFs) \cite{SP12,CA12}, of
which the usual posterior PMF is a specific member and the FM bound the corresponding bound.

In Sec. VII, we describe a specific application of a proposed detector that simultaneously localizes a point source
and performs its spectral characterization under daylight conditions. For this application,
we shall compare all of our newly derived bounds and the existing ones with the
numerically exact results for MI under varying operating conditions.

%
\section{Review} \label{sec:problem}

Let $\Theta$ be a discrete random input
taking values in the set $\Omega_\theta \defeq\{\theta_1,\ldots,\theta_M\}$
according to the PMF $p_m\defeq p_{\theta_m},\ m=1,\ldots,M$.
Let $X$ be a random output variable, {\it e.g.,} a measurement variable, that carries information
about $\Theta$, and is distributed according to the conditional probability density function (PDF),
$P(x|\theta_m)$, given the value $\theta_m$ of the input variable $\Theta$.
The output $X$ is in general an $N$-dimensional random vector taking values in some subset of $\mathbb{R}^N$.
If the $M$ different values that $\Theta$ can take may be regarded as representing $M$ different hypotheses,
then the Bayesian estimation problem of determining the value of $\Theta$
becomes the Bayesian multi-hypothesis testing problem.


The MI is the difference between the input entropy, $H(\Theta)$, and EE,
$H(\Theta|X)$,
\be
\label{e2}
I(\Theta;X)= H(\Theta)-H(\Theta|X),
\ee
where the two entropies on the right-hand side (RHS) are defined as
\begin{align}
\label{e3}
H(\Theta)&= -\sum_{m=1}^M p_m\, \log p_m;\nonumber\\
H(\Theta|X)&= -\bE_X\left[\sum_{m=1}^M p(\theta_m|X)\, \log p(\theta_m|X)\right],
\end{align}
where the symbol $\bE_X$ denotes averaging over the statistical fluctuations of $X$.
We shall omit the subscript when a joint expectation over both the input and output variables
is understood. Further, the base of the logarithm will be left unspecified throught this paper,
but whenever the logarithm is inverted, as in Sec.~IIB, the base $e$ of the natural logarithm,
denoted by ln, will be assumed.

\subsection{The Fano Bound}
Since $H(\Theta|X)$ may be regarded as the loss of information in the
passage through the communication channel, it is expected to be related to the
probability of error (PE) in estimating the input variable, $\Theta$. This is quantified
by the Fano upper bound on EE in terms of the PE, namely
\begin{equation}
\label{e4}
H(\Theta|X) \leq H(P_e)+P_e\log (M -1),
\end{equation}
or equivalently by the following lower bound on the successfully transmitted information:
\begin{equation}
\label{e5}
I(\Theta;X) \geq H(\Theta) - [H(P_e)+P_e\log (M -1)].
\end{equation}

\subsection{Feder-Merhav and Related Upper Bounds on the MPE}

The maximum {\it a posteriori} probability (MAP) decision rule yields the maximum
probability of making the correct detection among $M$ different hypotheses
based on observations. The minimum probability of error (MPE) in the $M$-detection
problem is one minus the former, namely
\be
\label{e6p}
P_e^{(min)} = 1-\bE_X[\maxm \{p(\theta_m|X)\}].
\ee

A tight upper bound on the MPE involving the generalized equivocation
entropy (GEE) was derived in an earlier paper \cite{SP12}, namely
\ba
\label{e19}
P_e^{(min)} &= 1-\bE_x\left\{\exp\left[-\int_1^\infty {dn\over n^2} H_n(\Theta|X=x)\right]\right\}
\nonumber\\
&\leq 1-\exp\left[-\int_1^\infty {dn\over n^2} \bE_x H_n(\Theta|X=x)\right]\nonumber\\
&= 1-\exp\left[-\int_1^\infty {dn\over n^2} H_n(\Theta|X)\right],
\end{align}
where $H_n(\Theta|X)$ denotes the GEE of order $n$ defined as
\be
\label{e20}
H_n(\Theta|X) = -\bE \ln p_n(\Theta|X), \ \ p_n(\theta_m|x) \defeq {p^n(\theta_m|x)\over \sum_m p^n
(\theta_m|x)}.
\ee
Use was made of the convexity of exponentiation to arrive at (\ref{e19}).
Note that for a given $x$, as $n$ increases $p_n(\theta_m|x)$ approaches 1
from below for $m=m_*(x)={\rm argmax}_m \{p(\theta_m|x)\}$ and 0 from above for
all other $m$ values. This increasing purification of the posterior $p_n(\Theta|X)$, also known as the tilted posterior PMF
of order $n$ \cite{CA12}, with increasing $n$ also implies a decreasing value of the GEE.

Since $H_1(\Theta|X) \geq H_n(\Theta|X)$ for all $n\geq 1$, it follows that
\ba
\label{e21}
P_e^{(min)} &\leq 1-\exp\left[-\int_1^\infty {dn\over n^2} H_1(\Theta|X)\right]\nonumber\\
&\leq 1-\exp[-H(\Theta|X)],
\end{align}
since the GEE of order 1 is the ordinary EE, $H(\Theta|X)$.
Transposing the exponential to one side and then taking the logarithm of both sides
of the resulting equation yields the FM bound on $H(\Theta|X)$:
\be
\label{e21a}
H(\Theta|X)\geq -\ln (1-P_e^{(min)}).
\ee

A tighter bound on the MPE results if we upper-bound the integral in the exponent in inequality (\ref{e19})
over two or more intervals, e.g.,
\ba
\label{e22}
\int_1^\infty {dn\over n^2} H_n(\Theta|X) &\leq \int_1^\lambda {dn\over n^2} H_1(\Theta|X)
+\int_\lambda^\infty {dn\over n^2} H_\lambda (\Theta|X)\nonumber\\
&=\left(1-{1\over\lambda}\right)H(\Theta|X) +{1\over \lambda} H_\lambda (\Theta|X),
\end{align}
which is smaller than $H(\Theta|X)$ used in the FM bound for any $\lambda > 1$.
For $\lambda=2$, for example, the following tighter upper bound on the MPE results:
\be
\label{e23}
P_e^{(min)} \leq 1-\exp\{-[H(\Theta|X)+H_2(\Theta|X)]/2\}.
\ee
Evidently many similar but still tighter bounds on the MPE involving more
and more of the higher-order GEEs may thus be derived,
but the advantage of the looser FM bound (\ref{e21}) is its immediate
relation to the more easily interpretable loss of statistical information
represented by EE.

\section{Other Derivations of the FM Bound and New Upper Bounds on MI and MPE}

An alternative derivation
of the FM bound follows immediately on noting that since the logarithm is a monotonically increasing function,
$H(\Theta|X)$, which is simply $-\bE\log p(\Theta|X)$,
is bounded below by $-\bE\log \maxm p(\theta_m|X)$. But
since the logarithm is also a concave function, the EE is less tightly bounded below as
\be
\label{e28}
H(\Theta|X)\geq -\log \bE_X \maxm p(\theta_m|X) = -\log(1-P_e^{(min)}),
\ee
which is the FM bound.

In preparing to generalize the FM bound, we present one other
derivation of it. If $\{s_1,\ldots,s_M\}$ is a PMF, {\it i.e.,} $\sum_m s_m=1,\ s_m\geq 0, \ \forall m$, then
\be
\label{e23a}
\max(a_1,a_2,\ldots,a_M)\geq \prod_{m=1}^M a_m^{s_m}.
\ee
It then follows from definition (\ref{e6p}) and the concavity of the logarithm that
\ba
\label{e23b}
\log(1-P_e^{(min)}) &\geq \log \bE \prod_{m=1}^M p^{s_m}(\theta_m|X)\nonumber\\
                    &\geq  \bE \left[\sum_m s_m(X) \log p(\theta_m|X)\right]\nonumber\\
                    &= B(s),
\end{align}
where the bound $B(s)$ is defined as
\be
\label{e23bb}
B(s)\defeq \bE \left[\sum_m s_m(X) \log p(\theta_m|X)\right].
\ee
and the non-negative powers $s_m$ have been allowed, for maximum generality, to depend on the output variable $X$,
subject to their sum being 1.

For the special choice, $s_m(X)=p(\theta_m|X)$, the RHS of inequality (\ref{e23b})
reduces to $-H(\Theta|X)$, which is the FM bound. A more general choice
involves the generalized posterior PMF, $p_n(\theta_m|X)$, defined earlier in (\ref{e20}),
that becomes sharper the larger its order $n$,
\ba
\label{e23c}
s_m(X) &= p_n(\theta_m|X) \nonumber\\
       &={p^n(\theta_m|X)\over Z_n(X)},\ \ m=1,\ldots, M,
\end{align}
where the partition function $Z_n$ is defined as
\be
\label{e23cc}
Z_n(X)\defeq \sum_{m=1}^M p^n(\theta_m|X).
\ee
To see that for this choice of the $s_m$, the lower bound $B$ defined by (\ref{e23bb}) becomes tighter with increasing $n$,
we first substitute (\ref{e23c}) into (\ref{e23bb}) and then calculate its derivative with respect to $n$,
\ba
\label{e23d}
{dB_n\over dn} = &(\log e)\, \bE_X\left\{ {1\over Z_n(X)}\sum_m p^n(\theta_m|X) \ln^2 p(\theta_m|X)\right.\nonumber\\
               &\left.-{1\over Z^2_n(X)}\left[\sum_m p^n(\theta_m|X)\ln p(\theta_m|X)\right]^2
               \right\}\nonumber\\
             = &(\log e)\,\bE_X\left[\langle \ln^2p(\Theta|X)\rangle_n-\langle \ln p(\Theta|X)\rangle_n^2\right],
\end{align}
where the triangular brackets with subscript $n$ indicate the expectation of the quantity they
enclose with respect to the generalized posterior $p_n$ defined by (\ref{e23c}).
Since the square is a convex function, the RHS of the last expression in (\ref{e23d})
cannot be negative, which means that $B$ is a non-decreasing function of $n$. Thus for $n>1$,
the bound $B$ is tighter than the FM bound, and represents an increasingly tighter sequence
of bounds on the MPE with increasing $n$. For $n\to\infty$, the generalized posterior $p_\infty(\theta_m|X)$
takes the value 1 for that single value of index $m$ for which $p(\theta_m|X)$
is the largest for all $m$ and the value 0 for all other values of $m$.
In other words, the bound $B$ in this limit may be expressed as
\ba
\label{e23e}
B_\infty = &\sum_m\int_{\cR_m} \log p(\theta_m|x) \, P(x)\, dx\nonumber\\
         = &\bE_X \log \max_m p(\theta_m|X),
\end{align}
where $\cR_m$ is the $m$th decision region chosen according to the MAP criterion that yields
the least M-hypothesis testing error probability.

A simple improvement of the FM bound is obtained on writing
\ba
\label{e32}
-\log p(\Theta|X) = &-\log \maxm p(\theta_m|X)\nn
                    &- \log \left[ p(\Theta|X)/\maxm p(\theta_m|X)\right],
\end{align}
and then calculating its expectation. In the second term on the RHS above, we perform this expectation 
first over the posterior and then use the convexity of the negative logarithm to reverse the order
of the logarithm and the posterior averaging to lower-bound that term before performing its expectation over the data $X$. 
A similar reversal of the order of the logarithm and expectation in the first term on the RHS of (\ref{e32})
also lower-bounds it. We thus obtain the following expectation over data as a lower bound on EE:
\be
\label{e33}
H(\Theta|X) \geq -\log (1-\MPE) +\Delta,
\ee
where $\Delta$ denotes the expression
\be
\label{e34}
\Delta\defeq - \bE_X\log\left[{\sum_m p^2(\theta_m|X)\over \maxm
p(\theta_m|X)}\right].
\ee
Note that the sum in expression (\ref{e34})
can be bounded as
\ba
\label{e35}
\minm p(\theta_m|X) \sum_m p(\theta_m|X)&\leq \sum_m p^2(\theta_m|X)\nn
                                        &\leq \maxm p(\theta_m|X) \sum_m p(\theta_m|X),
\end{align}
but since $\sum_m p(\theta_m|X)=1$ for any $X$, we have the bounds
\be
\label{e36}
\minm p(\theta_m|X) \leq \sum_m p^2(\theta_m|X) \leq \maxm p(\theta_m|X).
\ee
These bounds imply the following bounds on $\Delta$:
\be
\label{e37}
0\leq \Delta \leq \bE_X\log\left[{\maxm p(\theta_m|X)\over \minm
p(\theta_m|X)}\right].
\ee
Based on these bounds, we see that the lower bound (\ref{e33}) is tighter
than the first term on its RHS, which is the FM bound.

Due to the convacity of the logarithm, $\Delta$ as defined in (\ref{e34}) is lower bounded by
\ba
\label{e37a}
\Delta \geq -\log \bE_X &\Bigg[ p(\theta_{m_*(X)}|X)\nn
&+\sum_{m\neq m_*(X)} {p(\theta_m|X)\over p(\theta_{m_*(X)}|X)}
p(\theta_m|X)\Bigg],
\end{align}
where $m_*(X)$, defined as
\be
\label{e27}
m_*(X)=\argmax{m} p(\theta_m|X),
\ee
is the MAP estimator.

Since the ratio of the posterior probabilities
inside the square brackects in the above bound may be recast, via Bayes rule, as the ratio of the corresponding
joint PDFs while the first term inside those brackets upon taking its expectation is simply $1-P_e^{(min)}$
according to relation (\ref{e6p}), we may express (\ref{e37a}) as
\be
\label{e37b}
\Delta \geq -\log\left[1-P_e^{(min)}+\int dx \sum_{m\neq m_*(x)} {p_m^2\over p_{m_*(x)}}{P^2(x|\theta_m)
\over P(x|\theta_{m_*(x)})}\right].
\ee
In this form, the quantity $\Delta$ is lower bounded by an expression that can be evaluated via
integrals involving the conditional data PDFs alone. This is computationally efficient,
particularly if such PDFs are Gaussian.

Expression (\ref{e37b}) may be recast into a simpler-looking form by splitting the integration over $X$
into partial integrations over individual decision regions, $\cR_m$, $m=1,\ldots M$, chosen according to the MAP criterion.
The overall lower bound on the EE, tighter than the FM lower bound, now follows from (\ref{e33}),
\ba
\label{e37d}
H&(\Theta|X)\geq -\log\left[1-P_e^{(min)}\right]\nn
                & -\log\left[1-P_e^{(min)}+
                 \sum_m\int_{\cR_m} dx \sum_{m^\prime\neq m} {p_{m^\prime}^2\over p_m}{P^2(x|\theta_{m^\prime})
                 \over P(x|\theta_m)}\right],
\end{align}
in which the first term on the RHS is the FM lower bound.
A computationally more expedient but entirely equivalent form of the bound (\ref{e37d}) is the following:
\ba
\label{e37e}
H(\Theta|X)\geq &-\log\left[1-P_e^{(min)}\right]\nn
                & -\log\left[
                 \sum_m\int_{\cR_m} dx \sum_{m^\prime} {p_{m^\prime}^2\over p_m}{P^2(x|\theta_{m^\prime})
                 \over P(x|\theta_m)}\right].
\end{align}
Subtracting the bound (\ref{e37e}) from $H(\Theta)$ yields
the corresponding upper bound on MI.

For the case of a nonuniform prior in the low-SNR limit, the FM lower bound on EE is not tight,
since $1-P_e^{(min)}$ tends in this limit to the largest of the prior probabilities,
$p^>\defeq\maxm p_m$, and thus the FM bound tends to $-\log p^>$,
while the EE tends to the sum $-\sum_m p_m\log p_m$, which can be quite different.
By contrast, the improved lower bound (\ref{e37e}) is tighter for a nonuniform
prior, thanks to the second term on its RHS.

\subsection{A Useful Form for the Uniform Prior}

The concavity of the logarithm allows us to lower-bound EE as
\ba
\label{e39}
H(\Theta|X)&=-\bE \log p(\Theta|X)\nn
           &\geq -\log \bE p(\Theta|X)\nn
           &=-\log \bE_X \sum_{m} p^2(\theta_m|X).
\end{align}
Denoting the difference of the posterior from the prior as $\delta p(\Theta|X)=p(\Theta|X)-
p(\Theta)$, we may write the sum in (\ref{e39}) as
\ba
\label{e38}
\sum_m &[p_m + \delta p(\theta_m|X)]^2\nonumber\\
       &= \sum_m \left[p^2_m +\delta p^2(\theta_m|X)+2 p_m\delta p(\theta_m|X)\right]\nn
       &= {1\over M} +\sum_m\delta p^2(\theta_m|X),
\end{align}
in which we used the fact that, for the uniform prior, $p_m=1/M,$ $\forall m$, and
the last term on the RHS of the second line
vanishes, since the posterior, like the prior,
is a probability distribution that sums to 1. In view of (\ref{e38}),
the lower bound (\ref{e39}) may be re-expressed in terms of $\delta p(\Theta|X)$ as
\be
\label{e38a}
H(\Theta|X) \geq \log M -\log \left[1+{1\over M}\bE_X \sum_{m} \delta p^2(\theta_m|X)\right],
\ee
where the sum over $m$ thus differs from the corresponding sum over the prior probabilities, namely
$1/M$, by terms that are of quadratic order in the deviations $\delta p(\theta_m|X)$. This allows us to
upper-bound MI as
\be
\label{e39a}
I(\Theta;X) \leq \log \left[1+M\bE_X\sum_m\delta p^2(\theta_m|X)\right].
\ee

The MI bound (\ref{e39a}) is expected to
interpolate tightly between the $\log M$ value of MI in the high-discrimination
regime where only one of the posterior probabilities is close to 1 and all the other $(M-1)$ ones are close to 0,
and the vanishing value of MI in the low-discrimination regime where the correction contained
in the sum on the RHS in the upper bound (\ref{e39a}) tends to be small.
This correction term may be interpreted in terms of
an effective SNR, defined as the ratio
$\sum_m\bE_X\delta p^2(\theta_m|X)/\sum_m p_m^2$, in a manner rather analogous to
the SNR dependence of the information capacity of an additive Gaussian channel.

\section{A Sequence of Tighter Lower Bounds on Equivocation}

From the lower bound (\ref{e39}), we may derive a sequence of ever tighter bounds that approach it
while being interpretable in terms of a corresponding sequence of supplementary measures
of Bayesian error. We first write $p^2(\theta_m|X)$ as $p(\theta_m|X)[p(\theta_{m_*(X)}|X)+p(\theta_m|X)
-p(\theta_{m_*(X)}|X)]$, which, in view of the normalization of the posterior PMF, permits us to
write its sum as
\ba
\label{e42a}
\sum_m &p^2(\theta_m|X) = p(\theta_{m_*(X)}|X)\nn
  &- \sum_{m\neq m_*(X)}p(\theta_m|X) [p(\theta_{m_*(X)}|X)-p(\theta_m|X)],
\end{align}
where $m_*(X)$ is defined by (\ref{e27}). For a given value of data $X$, the first term is the likelihood
of deciding in favor of the hypothesis satisfying the MAP criterion, while the second term
is the reduction of that likelihood based on the incorrectness of the MAP based choice. That likelihood
reduction may be lower-bounded, in turn, by its minimum value, namely that given by replacing
$p(\theta_m|X)$ inside the square brackets in (\ref{e42a}) by its maximum value for $m\neq m_*(X)$,
which we denote by $p(\theta_{m_{**}(X)}|X)$. Since from normalization the sum of $p(\theta_m|X)$ over $m\neq m_*(X)$ is
simply $1-p(\theta_{m_*(X)}|X)$, the following upper bound on the sum (\ref{e42a}) is obtained:
\ba
\label{e42b}
\sum_m p^2(\theta_m|X) \leq  &p(\theta_{m_*(X)}|X) - [1-p(\theta_{m_*(X)}|X)]\nn
       &\times[p(\theta_{m_*(X)}|X)-p(\theta_{m_{**}(X)}|X)],
\end{align}
where the quantity $p(\theta_{m_{**}(X)}|X)$ is the second largest posterior probability for the given data value $X$.

Introducing the third largest posterior probability via a similar mathematical step tightens the
bound (\ref{e42a}) further. Specifically, the difference between the RHSs of (\ref{e42a}) and (\ref{e42b})
is
\be
\label{e42c}
\sum_{m\neq m_*(X),m_{**}(X)} p(\theta_m|X) [p(\theta_{m_{**}(X)}|X)-p(\theta_m|X)],
\ee
which is lower-bounded by the sum obtained by replacing $p(\theta_m|X)$ inside the square brackets
by $p(\theta_{m_{***}(X)}|X)$, the third largest posterior probability for the given value of $X$.
Since the sum of $p(\theta_m|X)$ over all $m$ except $m_*(X)$ and $m_{**}(X)$
is $1-p(\theta_{m_*(X)}|X) - p(\theta_{m_{**}(X)}|X)$, this yields
the following tighter bound for the sum (\ref{e42b}):
\ba
\label{e42d}
\sum_m p^2(\theta_m|X) \leq  &p(\theta_{m_*(X)}|X) - [1-p(\theta_{m_*(X)}|X)]\nn
                              &\times[p(\theta_{m_*(X)}|X)-p(\theta_{m_{**}(X)}|X)]\nn
                             &-[1-p(\theta_{m_*(X)}|X)-p(\theta_{m_{**}(X)}|X)]\nn
                             &\times [p(\theta_{m_{**}(X)}|X)-p(\theta_{m_{***}(X)}|X)].
\end{align}
The successive lowering of the upper bound eventually toward the exact value can be performed
by a recursive use of this mathematical strategy, but any improvement of the bound tightness
is expected to decrease sharply with each such successive step.

Taking first the expectation of the sequence of bounds (\ref{e42b}), (\ref{e42d}),$\ldots$
with respect to $X$ and then the negative logarithm of the resulting expressions yields, via (\ref{e39}),
the corresponding sequence of lower bounds on EE,
\be
\label{e42e}
H(\Theta|X)\geq -\log \left[1-P_e^{(min)}-\sum_{i=1}^n\delta_i\right],\ \ n=1,\ldots,M-1,
\ee
where the non-negative quantities $\delta_i$ are defined as
\ba
\label{e42f}
\delta_i\defeq \bE_X &\left\{\left[1-p(\theta_{m_*(X)}|X)-\ldots -p(\theta_{m_{(i)*}(X)}|X)\right]\right.\nn
                             &\left. \times \left[p(\theta_{m_{(i)*}(X)}|X)- p(\theta_{m_{(i+1)*}(X)}|X)\right]\right\}
\end{align}
in which $m_{(i)*}(X)$ is just a shorthand for the $\theta$ label corresponding to the $i$th largest
posterior probability, given $X$. One may regard the correction $\sum_i\delta_i$ in (\ref{e42e}) to be a bound on the
increase of the probability of hypothesis-decision error beyond the MPE given by the
MAP protocol.

\section{Generalization of the Feder-Merhav Approach and Other Lower Bounds on Equivocation}

In their seminal paper \cite{FM94}, Feder and Merhav derived a different lower bound on EE
as a function of the MPE, namely
\be
\label{e43}
H(\Theta|X) \geq \varphi_*(P_e^{min}),
\ee
where the function $\varphi_*$ is the piecewise linear, convex function defined as
\be
\label{e44}
\varphi_*(u) = \left\{
\begin{array}{ll}
   a_1u+b_1 & 0\leq u\leq 1/2\\
   a_2(u-1/2)+b_2 & 1/2\leq u\leq 2/3\\
   \vdots &\vdots\\
   a_n \left(u-{n-1\over n}\right)+b_n  & {n-1\over n}\leq u\leq {n\over n+1}\\
   \vdots &\vdots
\end{array}
\right. ,
\ee
where $a_n=n(n+1)\log[(n+1)/n]$ and $b_n=\log n$, $n=1,\ldots,M-1$.
The corresponding upper bound on MI, along with the bound (\ref{e28}), provide the tightest available
upper bounds on MI involving the MPE alone.

The Feder-Merhav approach can be refined by lower-bounding the EE more tightly, subject to the knowledge of the
first two highest posterior probabilities, not just the highest one that relates directly to MPE via (\ref{e43}).
Let $\{p_*(x),p_{**}(x),p_1(x),\ldots,p_{M-2}(x)\}$
be the set of all posterior probabilities arranged in descending order, given $x$. Here
$p_*(x)$ and $p_{**}(x)$ denote, in simplified notation, what we have previously called $p(\theta_{m_*(x)}|x)$ and
$p(\theta_{m_{**}(x)}|x)$. These two are to be held fixed
but the other $(M-2)$ posterior probabilities, $p_1(x),\ldots,p_{M-2}(x)$, are to be varied
subject only to the following constraints:
\ba
\label{e45}
\sum_{m=1}^{M-2}& p_m(x) =1-p_*(x)-p_{**}(x),\nn
&p_*(x)\geq p_{**}(x)\geq p_1(x)\geq p_2(x)\geq\ldots\geq p_{M-2}(x)\geq 0,
\end{align}
when minimizing $H(\Theta|X=x)=-p_*(x)\log p_*(x) -p_{**}(x)
\log p_{**}(x)+\delta H(\Theta|x)$, or equivalently $\delta H(\Theta|x)$,
where $\delta H(\Theta|x)=-\sum_{m=1}^{M-2}p_m(x)\log p_m(x)$.
Since the entropy is a concave function, the constrained minimum of
$\delta H(\Theta|x)$
is obtained at a constraint boundary, and the Kuhn-Tucker conditions for constrained minimization are active.
These conditions further imply that the sum representing the minimum value of $\delta H(\Theta|x)$
have as few equal terms containing the largest probabilities allowed by these constraints.

Because of constraints (\ref{e45}), $p_{**}(x)$ is required to be bounded above both by $\pi_*(x)$ and $1-\pi_*(x)$,
and thus by the smaller
of the two, where $\pi_*(x) \defeq 1-p_*(x)$ is the estimate of MPE. This leads to two possible cases and their
sub-cases, depending on the value of $p_*$ and $p_{**}$ (we drop the argument $x$ of these functions whenever clarity is
not affected):
\begin{enumerate}[label=\Alph*.]
\item \underline{$\pi_*(x)\leq 1-\pi_*(x)$, {\it i.e.,} $1/2\leq p_*\leq 1$}
\begin{enumerate}[label=\roman*.]
\item $0\leq \pi_*-p_{**}\leq p_{**}$, {\it i.e.,} $\pi_*/2\leq p_{**}\leq \pi_*$

For this subcase, the minimizing posterior PMF is $\{ p_*,p_{**}, \pi_*-p_{**},0,\ldots,0\}$, and the minimum
estimate of EE is
\ba
\label{e46a}
\min H(\Theta|x) =&-p_*(x)\log p_*(x)-p_{**}(x)\log p_{**}(x)\nn
                 &-(\pi_*-p_{**}(x))\log (\pi_*-p_{**}(x))\nn
                 =&h(p_*,p_{**}),
\end{align}
where $h(p,q)\defeq -p\log p-q\log q -(1-p-q)\log(1-p-q)$ is the ternary entropy for the PMF $\{p,q,1-p-q\}$.

\item $p_{**}\leq \pi_*-p_{**}\leq 2p_{**}$, {\it i.e.,} $\pi_*/3\leq p_{**}\leq \pi_*/2$

The minimizing posterior PMF is $\{ p_*,p_{**}, p_{**},\pi_*-2p_{**},0,\ldots,0\}$, and the minimum
possible estimate of EE is
\ba
\label{e46b}
\min H(\Theta|x) =&-p_*(x)\log p_*(x)-2p_{**}(x)\log p_{**}(x)\nn
                  &-(\pi_*-2p_{**}(x))\log (\pi_*-2p_{**}(x))\nn
                 =&2p_{**}\log 2+h(p_*,2p_{**}).
\end{align}

\item $2p_{**}\leq \pi_*-p_{**}\leq 3p_{**}$, {\it i.e.,} $\pi_*/4\leq p_{**}\leq \pi_*/3$

The minimizing posterior PMF is $\{ p_*,p_{**}, p_{**},p_{**},\pi_*-3p_{**},0,\ldots,0\}$, and the minimum
possible estimate of EE is
\ba
\label{e46c}
\min H(\Theta|x) =&-p_*(x)\log p_*(x)-3p_{**}(x)\log p_{**}(x)\nn
                  & -(\pi_*-3p_{**}(x))\log (\pi_*-3p_{**}(x))\nn
                 =&3p_{**}\log 3+h(p_*,3p_{**}).
\end{align}
\item And so on...
\end{enumerate}

\item \underline{$\pi_*(x)\geq 1-\pi_*(x)$, {\it i.e.,} $0\leq p_*\leq 1/2$}
\begin{enumerate}[label=\roman*.]
\item $0\leq \pi_*-p_{**}\leq p_{**}$, {\it i.e.,} $\pi_*/2\leq p_{**}\leq \pi_*$

Since $p_{**}\leq \min(\pi_*,1-\pi_*)
=1-\pi_*$, the bounds here can be tightened as $\pi_*/2\leq p_{**}\leq 1-\pi_*$, which requires for self-consistency
that $1/2\leq \pi_*<2/3$.
The minimizing posterior PMF is $\{ p_*,p_{**}, \pi_*-p_{**},0,\ldots,0\}$, and the minimum
possible estimate of EE is given by (\ref{e46a}).

\item $p_{**}\leq \pi_*-p_{**}\leq 2p_{**}$, {\it i.e.,} $\pi_*/3\leq p_{**}\leq \min(1-\pi_*, \pi_*/2)$

When $1-\pi_*\leq \pi_*/2$, the allowed range of $\pi_*$ is restricted to $2/3\leq \pi_*\leq 3/4$, but otherwise
$\pi_*$ need simply be bounded in $[1/2,2/3]$.
The minimizing posterior PMF is $\{ p_*,p_{**}, p_{**},\pi_*-2p_{**},0,\ldots,0\}$, and the minimum
possible estimate of EE is again given by (\ref{e46b}).

\item $2p_{**}\leq \pi_*-p_{**}\leq 3p_{**}$, {\it i.e.,} $\pi_*/4\leq p_{**}\leq \min(1-\pi_*,\pi_*/3)$

Again,
if $1-\pi_*\leq \pi_*/3$, the allowed range is restricted to the smaller interval $3/4\leq \pi_*\leq 4/5$, but otherwise
to the interval $1/2\leq \pi_*\leq 3/4$. In either case, the
the minimizing posterior PMF is $\{ p_*,p_{**}, p_{**},p_{**},\pi_*-3p_{**},0,\ldots,0\}$, and the minimum
possible estimate of EE is given by (\ref{e46c}).
\item And so on...
\end{enumerate}
\end{enumerate}

The results described above may be consolidated into the following
lower bound on the EE estimate, given the two highest posterior probabilities $p_*(x)$ and $p_{**}(x)$:
\be
\label{e47}
H(\Theta|X=x) \geq \varphi(p_*(x),p_{**}(x)),
\ee
where the function $\varphi$ takes different forms in the different allowed sub-reqions of the
$p_*-p_{**}$ plane as
\be
\label{e48}
\varphi(u,v) = \left\{
\begin{array}{ll}
   h(u,v)& {1-u\over 2}\leq v\leq \min(u,1-u)\\
   h(u,2v)+2v\log 2 & {1-u\over 3}\leq v\leq \min\left(u,{1-u\over 2}\right)\\
   \vdots &\vdots\\
   h(u,nv)+nv\log n  & {1-u\over n+1}\leq v\leq \min\left(u,{1-u\over n}\right)\\
   \vdots &\vdots
\end{array}
\right. .
\ee
We display these different allowed sub-regions in Fig.~1.
The full allowed region in the $p_*-p_{**}$ plane is the interior of the triangle OAB that bounds
$p_{**}$ as $0\leq p_{**}\leq \min(\pi_*,1-\pi_*)$. Over its triangular sub-regions ABC, ACD,
ADE, $\ldots$ designated by the upper-case
roman numerals I, II, III, $\ldots$ the different lower bounds
on the EE estimate starting from the top in (\ref{e48}) apply, respectively.
\begin{figure}[!ht]
\centering
\includegraphics[width=3in]{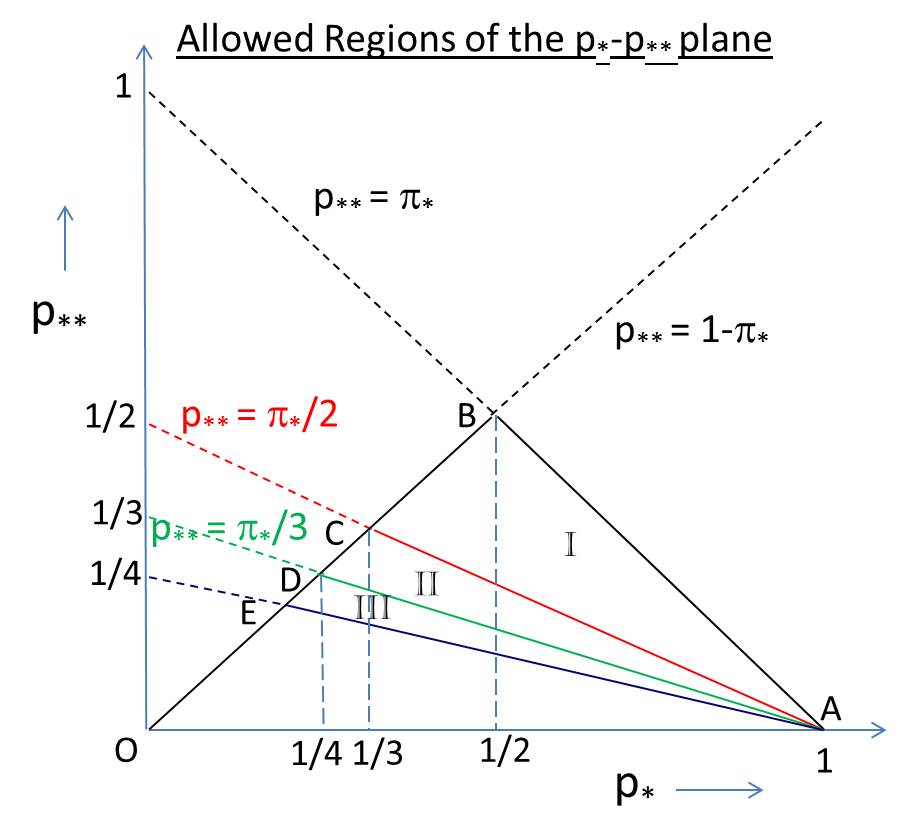}
\caption{\label{fig:f1} The overall allowed triangular region, OAB, in the $p_*-p_{**}$ plane and its various
sub-regions, ABC, ACD, ADE, etc, over each of which a separate form of the minimum value of the EE estimate,
subject to the given pair of values of $p_*$, $p_{**}$, applies. The text provides more detailed information.}
\end{figure}

Averaging the minimum value of the estimate of EE given by (\ref{e47}) and (\ref{e48})
over the output $X$ yields the lower bound on EE.
But since the ternary entropy is a concave function of its arguments, the tightest possible
convex lower bound on EE, which we denote as $\Phi(u,v)$, is a piecewise linear function of $u\defeq\bE_X p_*(X)$,
which is $1-\MPE$, and $v\defeq\bE_X p_{**}(X)$, with a separate linear form for each triangular sub-region
in the $p_*-p_{**}$ plane. For each such sub-region, if the function
$\varphi$ given in (\ref{e48}) takes values $\varphi_1$, $\varphi_2$, and $\varphi_3$ at its three vertices
whose coordinates are $(u_1,v_1)$, $(u_2,v_2)$, and $(u_3,v_3)$, respectively, then
$\Phi(u,v)$ in the interior of the triangle at a point $(u,v) = \sum_{i=1}^3 (u_i,v_i)\, t_i$, with
$t_i\geq 0, \ \sum_{i=1}^3t_i=1$, is equal to the value of the corresponding linear interpolant at that point,
$\Phi(u,v)=\sum_{i=1}^3 t_i\varphi_i$.
With the following values of $\varphi$ at the various triangular vertices in Fig.~1, as simply obtained from (\ref{e48}):
\ba
\label{e49}
\varphi_A&=h(1,0)=0;\ \varphi_B=h\left({1\over 2},{1\over2}\right)=\log 2; \nn
\varphi_C&=h\left({1\over3},{2\over3}\right)+ {2\over 3}\log 2;
\ \varphi_D=h\left({1\over4},{3\over4}\right)+{3\over 4}\log 4; \ldots;\nn
\varphi_n&=h\left({1\over n+1},{n\over n+1}\right)+{n\over n+1}\log n;\ldots,
\end{align}
where $\varphi_n$ is its value at $(1/(n+1),1/(n+1))$, the $n$th vertex point when going down the $45^0$ diagonal line,
the convex lower bound function $\Phi(u,v)$ is then fully specified in the allowed region, $v\leq \min(u,1-u)$, $u,v\geq 0$.
Since at these vertices the ternary entropy function reduces to the binary entropy function,
the convex lower bound $\Phi(u,v)$ involves only the binary entropy function in all allowed sub-regions of the $p_*-p_{**}$
plane.

The overall convexity of the so constructed piece-wise linear function $\Phi$ over the entire allowed region is
easily checked numerically by verifying the convexity of the function along the main diagonal
in the $uv$ plane, and is also clearly seen in Fig.~2 in which we display this function.
\begin{figure}[!ht]
\centering
\includegraphics[width=3in]{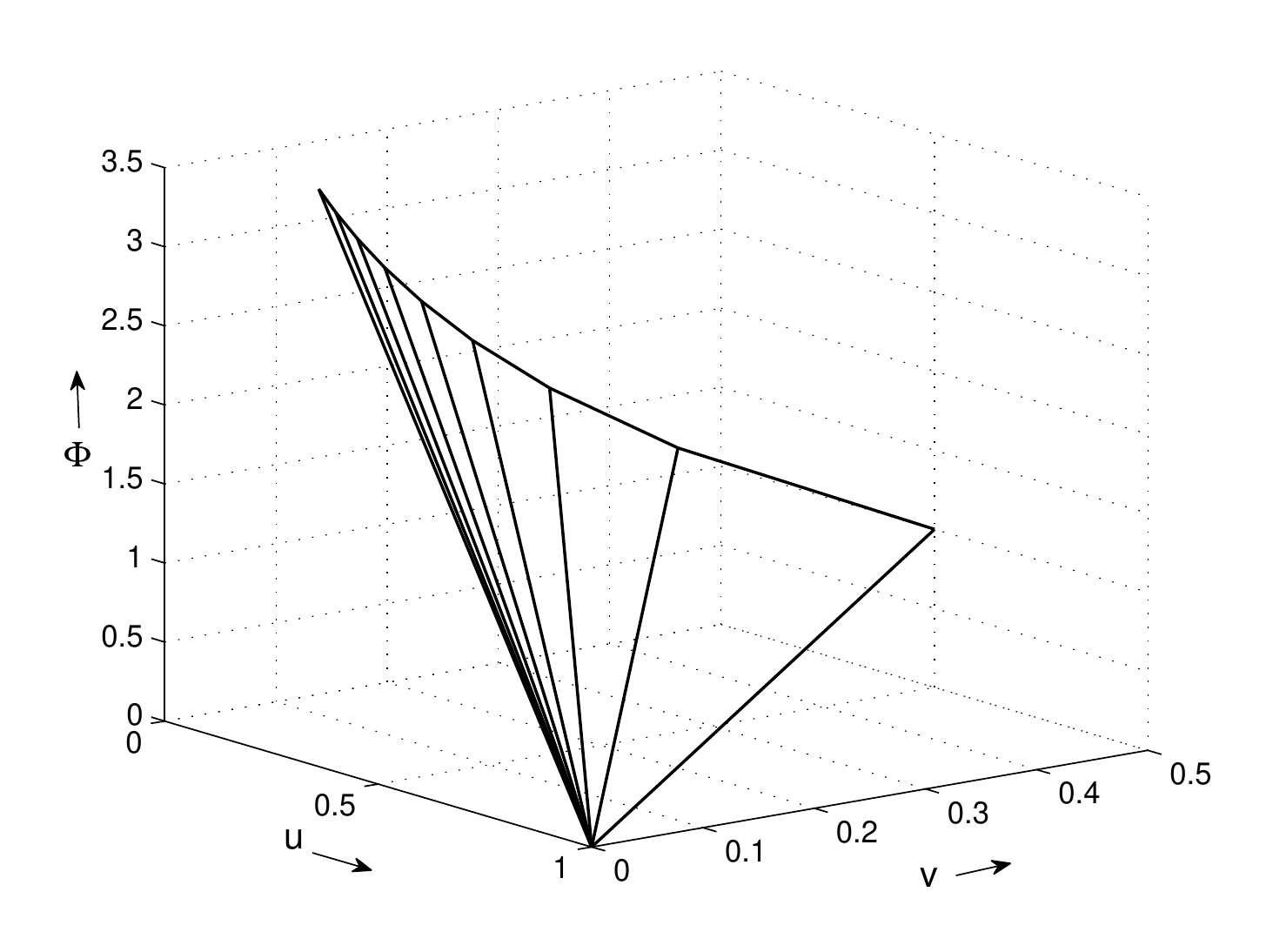}
\caption{\label{fig:f2} The piecewise linear convex bound function $\Phi(u,v)$}
\end{figure}

Further refinement of the piecewise linear convex FM bound can be
performed similarly as here, tightening the bound further,
by including increasingly more of the posterior probabilities
in the descending order of their values. We expect this procedure to improve the tightness of
the bound more rapidly at the higher SNR values than at the lower ones where more, rather than fewer,
of the so-ordered posterior probabilities are needed to specify the problem more accurately.

\section{An Improved Sequence of Upper Bounds on Equivocation}

As in the previous section, let us fix the two highest posterior probabilities, $p_*(x)$ and $p_{**}(x)$,
and maximize the estimator for equivocation, $H(\Theta|X=x)$,
subject to the sum of the remaining $(M-2)$ posterior probabilities being
$1-p_*(x)-p_{**}(x)$. The maximum is easily shown
to be attained when these remaining probabilities take equal values, each being
$[1-p_*(x)-p_{**}(x)]/(M-2)$, which yields the following upper bound
on the equivocation estimate, given the data value $x$:
\ba
\label{e50}
H(\Theta|&X=x) =-\sum_{m=1}^M p(\theta_m|x)\,\log p(\theta_m|x)\nn
             & \leq - p_*(x)\,\log p_*(x)  - p_{**}(x)\, \log p_{**}(x)\nn
             & - [1-p_*(x)-p_{**}(x)] \log\left[{1-p_*(x)-p_{**}(x)\over M-2}\right].
\end{align}
Taking the expectation of this bound with respect to $X$ and noting that
$-u\log u$ is a concave function, we may obtain the following looser upper
bound on equivocation in terms of the MPE and the largest contribution to it, namely
$\delta P_e^{(1)}\defeq \bE_X[p_{**}(X)]$:
\ba
\label{e51}
H(\Theta|X) & \leq - [1-\MPE]\,\log [1-\MPE] -\delta P_e^{(1)}\log\delta P_e^{(1)}\nn
             & - [\MPE-\delta P^{(1)}_e]\,\log [\MPE-\delta P_e^{(1)}] \nn
             &+ [\MPE-\delta P^{(1)}_e]\log (M-2).
\end{align}

Just like the lower bounds (\ref{e42e}) derived earlier, the upper bound (\ref{e51})
can be tightened further by fixing increasingly more of the posterior probabilities ordered according to
decreasing values. The upper bound (\ref{e51}) is tighter than the original Fano bound (\ref{e4}), which
can be regarded as the zeroth-order bound in our sequence of upper bounds, one that depends
only on the expectation of the highest posterior probability, which is simply $1-\MPE$, with the remaining $(M-1)$
probabilities chosen to be each equal to $\MPE/(M-1)$.

We shall now compare the tightness of the various lower and upper bounds on EE we have derived here.
The specific problem for which we do so is that of jointly estimating the
angular position and chemical signature of a muzzle flash using a proposed FLEM-FLASH system.

\section{Comparison of the Various Bounds for a Special Example}

The Fly's Eye Muzzle-Flash (FLEM-FLASH) detection system consists of a number of single-pixel micro-imagers,
essential simple light buckets, that are arranged
in a regular pattern on an opaque spherical surface. Each of the $S$ micro-imagers is capable of
multi-spectral measurements, say in $B$ spectral bands, but the signal photons
they record are limited both by the geometrical vignetting of the light-collection micro-tubes
in which the imagers are housed and the obliquity of illumination, which reduces the
the number of counts when compared to that for a head-on-illuminated imager by the factor $\cos\theta$.
The obliquity angle $\theta$ is the angle between the imager normal and the
direction from which the muzzle flash is incident. The multi-spectral counts are corrupted
both by a possibly bright but fluctuating background, as for day-time observations, and the
read-out noise of detection. A simple centroiding of the
spatial distribution of counts on the spherical surface can provide a good estimate of
the directionality of the flash, while a spectral pattern matching can reveal
the chemical signature of the flash. A schematic of a one-dimensional (1D) ring version
of the FLEM-FLASH system is shown in Fig.~3. Such a 1D version of the full 2D spherical
system suffices for the purposes of this paper.
The FLEM-FLASH system was proposed by the author as part of a DARPA funded project
under its SCENICC program in 2012 \cite{DARPA12}.

\begin{figure}[!ht]
\centering
\includegraphics[width=3in]{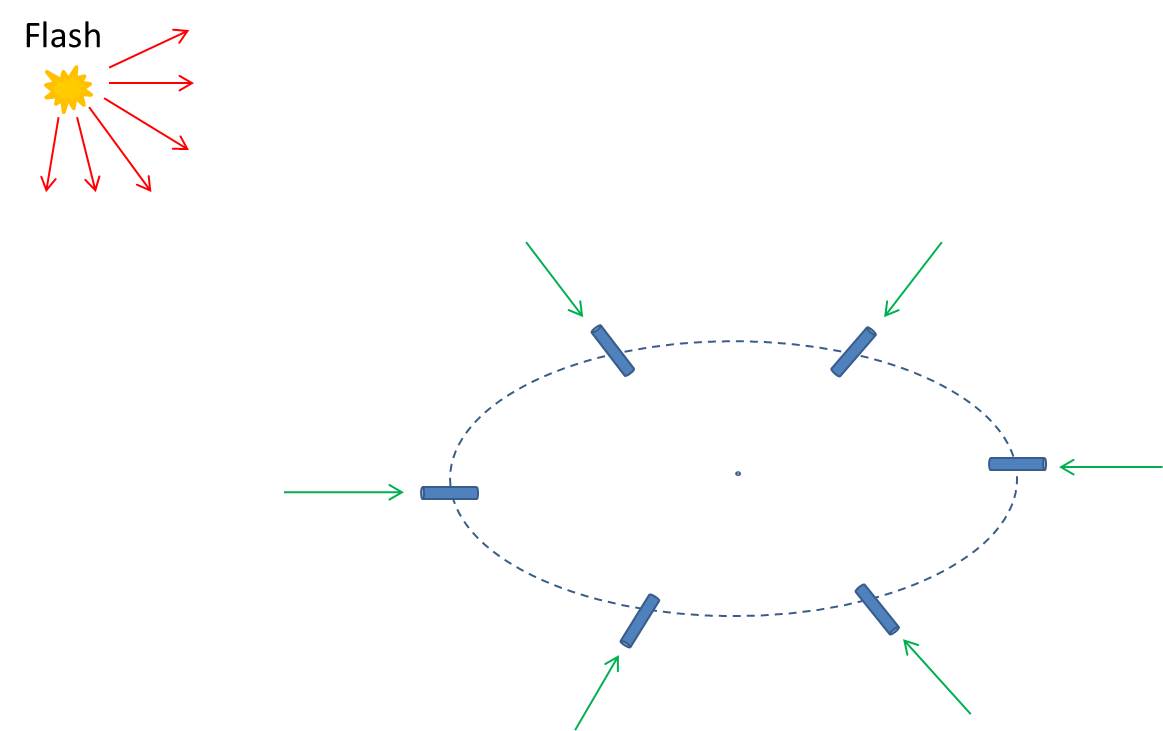}
\caption{\label{fig:f3} A schematic one-dimensional ring of regularly spaced single-pixel multi-spectral
micro-imagers. The imagers are oriented radially outwards, so each receives maximum head-on illumination
from a point source, like a muzzle flash, along the radial line on which it is located. These head-on
illumination directions for the different imagers are indicated by green arrows. }
\end{figure}

The MI between the detected multi-spectral
count data and the pair of input parameters, namely flash directionality and spectral signature,
yields an upper limit on the amount of statistical information about the input parameters that
the count data can provide to {\it any} estimator of these parameter values.
Specifically, $\Theta$ here is a two-component vector whose possible values range
over an outer product of 4 possible flash directions and 8 possible spectral signatures,
the latter distinguished by their spectral traces, {\it i.e.,} over 32 possible directionality-spectral
signature pairs. We first take the prior over these possible pair values to be uniform, so
$H(\Theta)=\log_2(32) = 5$ bits. The conditional fluctuations of the background and read-out noise are
assumed to be describable by Gaussian statistics, conditioned on the knowledge of
the value of the parameter vector, {\it i.e.,}
\be
\label{e31}
X|\Theta\sim {\cal N} (\underline{m}(\Theta),\sigma^2),
\ee
where $\sigma^2$ is the sum of the sensor read-noise variance, set here at 25 squared electrons, and
the variance of the background brightness fluctuations, considered isotropic and
uniform across the different spectral channels. The latter
fluctuations are taken to be a fraction, set here at 50\%, of the
mean background brightness level, set here at 10 electron counts. The ratio of the
mean signal $s$ for head-on illumination of a microimager and the mean background level $b$
defines the peak signal-to-background ratio (PSBR).

In Fig.~4, we plot an exact Monte-Carlo sampling based evaluation
of the exact fractional MI (fMI), which is simply the ratio of MI to $H(\Theta)$ that is bounded between
0 and 1, by means of the dashed curve for a number of values of the PSBR.
\begin{figure}
\centerline{
\includegraphics[width=3.5in]{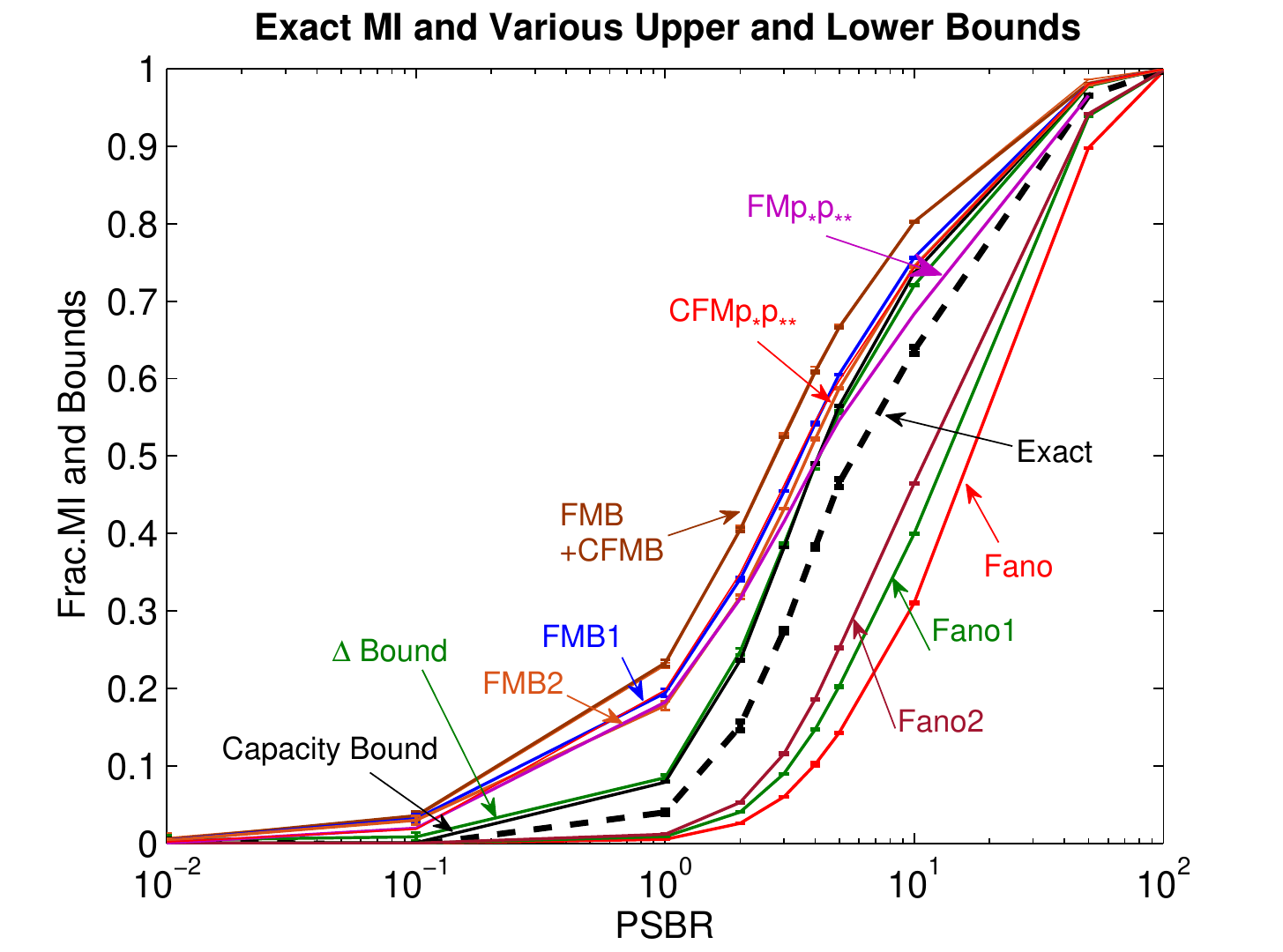}}
\caption{\label{fig:f4} Plot of the numerically evaluated MI and its various bounds vs. peak signal-to-background
ratio (PSBR), along with their relatively small errorbars. The color of the line segments through the various points on these plots
indicates the type of bound, as defined in the text.}
\end{figure}
All the other curves on this plot depict bounds on the exact fMI. The upper bounds on fMI are
given by subtracting the lower bounds on EE from $H(\Theta)$,
the lower bounds by subtracting the upper bounds on EE from $H(\Theta)$, and then normalizing
the difference in each case by $H(\Theta)$ itself.
Below we refer to the bounds on fMI as being related in this way to the corresponding bounds on EE derived
in this paper. We use the symbol FMB to denote the Feder-Merhav bound corresponding to (\ref{e28});
CFMB to denote the slightly tighter, minimally convex FM bound corresponding to (\ref{e43});
FMB1 and FMB2 to denote the first two improved FM bounds corresponding to (\ref{e42e}) for $n=1$ and $n=2$;
FMp$_*$p$_{**}$ to denote the refined FM bound corresponding to (\ref{e48});
CFMp$_*$p$_{**}$ to denote the refined but looser FM bound obtained by piece-wise linear
convexification of (\ref{e48});
capacity bound to denote bound (\ref{e39a}); $\Delta$-bound for the bound corresponding to (\ref{e37e});
Fano to denote the Fano lower bound (\ref{e5}); and Fano1 and Fano2 to denote, respectively, the first and second-order
corrections to the Fano bound corresponding to (\ref{e51}) and its next improved version involving $\MPE$, $\delta P_e^{(1)}$,
and the third largest posterior probability, $\delta P_e^{(2)}\defeq \bE_X[p(\theta_{m_{***}(X)}|X)]$.

Note that the FMB and CFMB are essentially indistinguishable, but including $\delta_1$ and
subsequently $\delta_2$ in (\ref{e42e}) to yield the first two corrections to the FM bound
tightens it significantly as indicated by the line segments labeled as FMB1 and FMB2.
The FMp$_*$p$_{**}$ bound, which is the refined FM bound without convexification, is
excellent at moderate to high PSBR values, but is no better than FMB2 at much lower PSBR.
This is expected since the lower the PSBR value the more mutually comparable values
all the posterior probabilities take, and so including just the highest two posteriors is inadequate
to accurately represent the exact result, but not so at the higher PSBR values.
The convexified bound, CFMp$_*$p$_{**}$, is considerably looser and virtually indistinguishable
from FMB1 at all PSBR values, but still tighter than
FMB which only includes the highest posterior probability.

The capacity bound (\ref{e39a}) is quite a bit tighter than these, however,
since the former represents the value of (\ref{e42e}) that would be obained were we to set $n=M-1=31$ there
and thus what we would label as FMB31
for the present problem. The $\Delta$-bound is nearly coincident with the capacity bound for the present
problem over the full PSBR range.
The two improved Fano lower bounds, labeled by Fano1 and Fano2, tighten the classic Fano bound
considerably, but the gap between the exact results for fMI and these tightened bounds is still quite significant
for the present application.

\section{Concluding Remarks}
In this paper we have derived a number of useful upper and lower bounds on MI
in terms of the associated MPE, which is directly related to the expectation of the maximum
posterior probability, and certain secondary Bayesian metrics of error
that involve the second, third, $\ldots$ largest of the posterior probabilities.
By incorporating these secondary error metrics, we are able to tighten the well known Feder-Merhav and
Fano bounds on MI that are based exclusively on the MPE alone.

We have also derived an upper bound on MI that is tighter still, but it does not have a
simple interpretation in terms of Bayesian error. For high-dimensional problems, however, it does provide a computationally
expedient, accurate approximation to MI that can be useful when exact MI calculations would be rather prohibitive.
All our bounds were compared to the numerically exact calculations of MI to assess their
tightness for a specific application.

We have also derived a family of upper bounds on the MPE itself defined in terms of the generalized, or tilted, posterior
PMF, which can be used, in a sequential manner, to tighten that presented by the EE in the Feder-Merhav bound.
While the bound is not attained by the MPE even as the minimum of the sequence for a finite-dimensional
problem, it is expected to be tight in the asymptotic limit of infinitely many data variables.
These results, as well as the bounds on MI involving Bayesian error probabilities that were derived in this paper,
can ultimately be useful for a study of error in channel coding and data compression when finite codewords and bit strings
are involved and the asymptotic limit is not particularly meaningful.

\section*{Acknowledgments}
The author is pleased to acknowledge helpful contributions from S. Narravula.
Funding support from the Air Force Office of Scientific Research under grants
FA9550-08-1-0151 and FA9550-09-1-0495 is gratefully acknowledged.

\end{document}